\documentclass[twocolumn,showpacs,preprintnumbers,amsmath,amssymb]{revtex4}
\usepackage{graphicx}% Include figure files
\usepackage{dcolumn}% Align table columns on decimal point
\usepackage{bm}% bold math
\newcommand {\ket}[2]{\bigl | {#1} \rangle_{#2}}
\newcommand {\bra}[2]{{_{#2}\langle} {#1}\bigr |}
\newcommand {\sclr}[2]{\langle \, #1 \bigr | #2 \, \rangle}
\newcommand {\mtr}[3]{\langle \, #1 \bigr | #2 \bigl | #3 \, \rangle}
\newcommand {\rmtr}[3]{\langle \, #1 \|  #2 \|  #3 \, \rangle}
\newcommand {\ve}{\varepsilon}
\newcommand {\vf}{\varphi}
\newcommand {\eqrf}[1]{Eq.(\ref{#1})}
\newcommand {\eqct}[1]{(\ref{#1})}
\newcommand {\fig}[1]{Fig.\ref{#1}}
\newcommand {\rf}[1]{Ref.\cite{#1}}
\newcommand {\ct}[1]{\cite{#1}}
\newcommand {\ham}{\hat{\mathcal H}}

\newcommand {\vc}[1]{{\bf #1}}
\newcommand {\mc}[1]{\mathcal #1}

\newcommand {\eqt}[2]{\begin{equation} #2 \label{#1} \end{equation}}
\newcommand {\pr}{^{\,\prime}}

\newcommand {\shr}{Schr\"{o}dinger }
\newcommand {\suml}{\sum\limits}
\newcommand {\intl}{\int\limits}
\newcommand {\re}{\mathcal{R}e}
\newcommand {\im}{\mathcal{I}m}
\newcommand {\ft}[1]{\footnotemark[#1]}
\begin{document}

\title{Generalized optical potential for weakly bound nuclei. \\ I. Two-cluster projectiles.}

\author{A.S. Denikin}%\email{denikin@jinr.ru}
\author{V.I. Zagrebaev}
\affiliation{Flerov Laboratory of Nuclear Reactions, JINR, Dubna, 141980 Russia}
\author{and P. Descouvemont}
\affiliation{Physique Nucl\'{e}aire Th\'{e}orique et Physique Math\'{e}matique, CP229 \\
ULB, B-1050 Brussels, Belgium}

%\date{\today}

\begin{abstract}
{
A generalized optical potential for elastic scattering induced by light nuclei is calculated within the Feshbach projection operator method. The model explicitly takes into account the contribution of the projectile break-up continuum treated within a microscopic few-cluster model. In this work we formulate the model, deriving an explicit expression for the optical potential, and show ability of the model applying it to deuteron elastic scattering.
}
\end{abstract}

\pacs{24.10.-i, 24.10.H, 25.45.-z, 25.60.Bx}

\maketitle

\section{Introduction}\label{intro}
During the last years reactions with light weakly bound nuclei have been of increased interest from the experimental and theoretical point of views and the progress in the investigation of these nuclei has been impressive \cite{jonson04, kelley07}. This progress is conditioned by the important efforts devoted to investigate reaction mechanisms and by new techniques enabling the production of exotic nuclei. In spite of that we are still far from a clear understanding of the unusual structure of exotic nuclei and of the reaction mechanisms induced by these nuclei. This arises from the experimental difficulties (also by low intensities of available beams) and by the difficulties arising in the description of a few-body nuclear dynamics taking place under conditions of a strong coupling of all reaction channels with the break-up channel of weakly bound projectiles.

Generally it is assumed that in light exotic nuclei the nucleons tend to group into clusters, whose relative motion mainly defines the properties of these nuclei. This assumption leads to great advantages for models employing the cluster concept both for the structure and reactions involving light exotic nuclei \cite{suzuki, zhukov93, danilin98, danilin06, descHH, zagreb99, ershov01, chatt01, rusek01, mackintosh04, abuibrahimPRC04, semkin95, soper}. In the case of nuclear reaction study the coupled-channel (CC) formalism is the one of consistent and efficient approaches \cite{rawitscher, sakuragi87, austern87, hagino99, zagreb04}. The continuum-discretized reduction of this method (CDCC) allows to study the interplay between elastic and break-up channels in reactions involving two-cluster nuclei \cite{soper, rawitscher, sakuragi87, austern87, mackintosh04, diaz02, rusek01}. However, the application of the CC approach to reactions with few-body projectiles meets significant computational difficulties if a realistic few-body wave function is used. Consequently very few works have been done \cite{matsumoto04, matsumoto06}.

The generalized optical model (GOM) of H. Feshbach \ct{feshbach} is an alternative approach to the problem. Models based on the Feshbach theory are extensively used for the study of coupling effects on different reaction channels. It is worth to mention, in particular, the studies of the role of the deuteron break-up in its elastic scattering with heavy ions \ct{testoni, baumg, mukherjee68, zagrebaev78}, the influence of collective excitations on heavy ion elastic scattering \ct{love, hussein84, khadkikar81, coulter77, vinhmau, derm, pacheco}, the interplay between break-up and complete fusion channels in weakly bound nucleus reactions \ct{hussein93}. However the application of the Feshbach method was hampered in the past by the complexity of the formulation and of computational burden. As a result, even in studies of the deuteron break-up, the applications of this method have been done with many simplifications.

In contrast with the CDCC approach, the GOM allows one to avoid the simplifying discretization of the continuum spectrum. The calculation within the GOM is faster. On the other hand, the CDCC method provides phase-shifts in all reaction channels, rather treats the continuum-to-continuum coupling, and does not require an approximation for the Green function, while the GOM does. Therefore in the case of a two-cluster projectile the CDCC approach is somewhat more efficient. However, in the case of a few-body projectile the application of the CDCC method becomes difficult (the large size of the coupling matrix, the complicate procedure of matrix element calculation and e.g.), while the GOM remains more feasible.

Our main goal is to apply the GOM to study the elastic scattering of light weakly bound nuclei using a realistic few-body model description of their internal structure. As a test of the approach, in the present work, the GOM is applied to the deuteron elastic scattering from heavy targets at intermediate energies in order to draw conclusions about its applicability. Applications of the model to reactions with few-cluster weakly bound nuclei (such as $^6$He) will be done in a subsequent publication.

The method proposed earlier \cite{testoni, zagrebaev78} is extended here in order to avoid simplifying assumptions. In \cite{testoni, zagrebaev78} the authors (i) neglected the spins of the particles, (ii) considered coupling with $s$-wave continuum only, (iii) neglected or treated the Coulomb forces in an approximate way, and (iv) used the free-particle Green function instead of the total one. Our approach goes beyond these assumptions. Within the method a structureless target nucleus interacts with a projectile treated as a system of few bound clusters. The bound and continuum states of the projectile are described in the framework of the microscopic cluster model and used to construct the Feshbach projection operators. We derive an explicit expression for the optical potential which takes into account explicitly the coupling with projectile break-up channels. We show also the importance of an accurate treatment of the Green function appearing in the dynamical polarization potential in the case of light targets.

%----------------------------------------------------------------------------------------------------------------------
\section{Model}\label{model}
We consider the scattering of a weakly bound projectile by a structureless target. The projectile is treated as a bound few-cluster system. The corresponding Hamiltonian has the following form
%+++++++++++++++++++++++++++++++++++++++++++++++Equation+++++++++++++++++++++++++++++
\begin{equation}
\label{ham} \ham_{\bf R,\xi} = \hat T_{\bf R} + \ham_{\bf \xi} + \hat V_{\bf R,\xi},
\end{equation}
%+++++++++++++++++++++++++++++++++++++++++++++++++++++++++++++++++++++++++++++++++++++++
where $\hat T_{\bf R}$ is the kinetic energy operator of the projectile-target relative motion, $\ham_{\bf \xi} = \hat t_{\bf \xi} + \hat v_{\bf \xi}$ is the Hamiltonian describing the projectile internal structure, and $\xi$ denotes an appropriate set of internal coordinates. The interaction potential $\hat V_{\bf R,\xi}$ is a sum of effective (non-hermitian) cluster-target potentials, which is obtained from a fit of elastic scattering data. By employing these cluster-target interactions we implicitly take into account the internal properties of the clusters and target.

The total scattering wave function $\ket{\Phi}{\vc R,\xi}$ satisfies the \shr equation with the Hamiltonian \eqct{ham} and eigenvalue $E = E_p + \ve_0$, where $E_p = \hbar^2p^2/2m$ is the projectile-target relative energy, $m$ is the reduced projectile-target mass, while $\ve_0$ is the projectile ground state energy. The ground $\ket{\ve_0,j_0}{\xi}$ (here $j_0$ is a projectile total angular momentum) and continuum $\ket{\kappa}{\xi}$ states of the projectile form together the spectrum of the Hamiltonian $\ham_{\xi}$, since the weakly bound projectile is supposed to have only one bound state. The Feshbach projection operators $\hat P_\xi$ and $\hat Q_\xi$ \cite{feshbach} are constructed as follows
%--------------------------------------------------------------------
$$\hat P_\xi + \hat Q_\xi = \ket{\ve_0,j_0}{}\bra{\ve_0,j_0}{} + \int \ket{\kappa}{}\bra{\kappa} {} \; d\kappa.$$
%--------------------------------------------------------------------
The operator $\hat P_\xi$ extracts the elastic component of the total wave function $\hat P_\xi \ket{\Phi_{\bf p}}{\bf R,\xi} = \ket{\Psi_{\bf p}}{\bf R,\xi}$, which satisfies the \shr equation
%+++++++++++++++++++++++++++++++++++++++++++++++Equation+++++++++++++++++++++++++++++
\begin{eqnarray}
&\Bigl( \hat T_{\bf R} + \bigl[ \hat P \hat V_{\bf R,\xi} \hat P + \hat P \hat V_{\bf R,\xi}\hat Q \frac{1}{E - \hat Q \ham_{\bf R,\xi} \hat Q}  \hat Q \hat V_{\bf R,\xi} \hat P \bigr]\Bigr) \ket{\Psi_{\bf p}}{\bf R,\xi} &\nonumber\\ & = (E-\ve_0) \ket{\Psi_{\bf p}}{\bf R,\xi}& \label{omp1}
\end{eqnarray}
%+++++++++++++++++++++++++++++++++++++++++++++++++++++++++++++++++++++++++++++++++++++++
where the expression in the square brackets is the generalized optical potential. The first term $\hat U^{(1)} = \hat P \hat V_{\bf R,\xi} \hat P$ is the local cluster-target interactions folded over the projectile ground state. The second term (we will refer to it as $\hat U^{(2)}$) is the non-local dynamical polarization potential (DPP), describing the coupling of elastic and non-elastic channels.

\subsection{Optical potential for $N$-cluster projectiles}\label{Nbody}
Following the usual technique \cite{SatchlerDNR} the elastic component of the many-body wave function may be expanded in partial waves as %+++++++++++++++++++++++++++++++++++++++++++++++++++++++++++++++++++++++++++++++++++++++
\begin{eqnarray}
& \Psi_{\vc p, j_0m_0}^{(+)}(\xi,\vc R) = \nonumber \\ & \frac1{pR} \suml_{JL} i^L e^{i\sigma_{L}} \sqrt{\frac{2L+1}{2\pi^2}}\, \;\psi_{JLj_0}(p,R)\,\Theta_{Jm_0}^{Lj_0}(\xi,\Omega_R), \label{el_wf3}
\end{eqnarray}
%+++++++++++++++++++++++++++++++++++++++++++++++++++++++++++++++++++++++++++++++++++++++
where $\vc p$ is supposed to be parallel to the $z$-axis. $\psi_{JLj_0}(p,R)$ is a partial wave function describing the projectile-target relative motion with total $(J)$ and orbital $(L)$ angular momenta, and with asymptotic
%---------------------------------------------------------------------------------------
\begin{equation}
\psi_{\nu}(p,R\to \infty) \to F_L(pR) + \frac{S_{\nu}-1}{2i} H_L^{(+)}(pR),\label{boundary}
\end{equation}
%---------------------------------------------------------------------------------------
where $S_{\nu} = e^{2i\delta_{\nu}}$ is a scattering S-matrix element, $H^{(+)}_L = G_L+iF_L$, while $F_L$ and $G_L$ are the regular and irregular Coulomb functions. In \eqrf{el_wf3} we use the notation $\Theta_{JM}^{Lj_0} (\xi,\Omega_R)$ for the spin-angle wave function resulting from the $\vc L$ and $\vc j_0$ vector coupling
%-----------------------------------------------------------------------------------------
\begin{equation}
\Theta_{JM}^{Lj_0}(\xi,\Omega_R) = \suml_{\gamma_0} \phi_{j_0\gamma_0}(\xi)\, \Bigl[ \mc Y_{j_0}^{\gamma_0}(\Omega_\xi) \otimes Y_L(\Omega_R) \Bigr]_{JM}, \label{THETA}
\end{equation}
%-----------------------------------------------------------------------------------------
where $\phi_{j_0\gamma_0}(\xi)\;\mc Y_{j_0 m_0}^{\gamma_0}(\Omega_\xi)$ is a partial component of the projectile ground state wave function.

We substitute expansion \eqct{el_wf3} in the \shr equation \eqct{omp1}. Multiplying the resulting equation by $\Theta_{J\pr M\pr}^{\dag\,L\pr j_0}(\vc r,\Omega_R)$ from the right, and integrating over $\vc r$ and $\Omega_R$,  one obtains a set of coupled \shr equations for the partial wave functions
%\psi_{\nu}^{''}(R) + +++++++++++++++++++++++++++++++++++++++++++++++++++++++++++++++++++++++++++++++++++++++
\begin{eqnarray}
& \Bigl( \frac{d^2}{dR^2} + p^2 - \frac{L(L+1)}{R^2}\Bigr) \psi_{\nu}(R) - \frac{2m}{\hbar^2} \suml_{L\pr} \Bigl(U_{LL\pr}^{(1)}(R)\, \psi_{\nu\pr}(R) \nonumber \\ & + R \intl_0^\infty R\pr U_{LL\pr}^{(2)}(R,R\pr)\, \psi_{\nu\pr}(R\pr)\, dR\pr \Bigr) = 0, \label{part_shr2}
\end{eqnarray}
%+++++++++++++++++++++++++++++++++++++++++++++++++++++++++++++++++++++++++++++++++++++++
where index $\nu$ denotes the set of quantum numbers $\{JLj_0\}$. The sets $\nu$ and $\nu\pr$ differ in orbital momentum $L$ and $L'$ only, and are used here just for the sake of simplicity. Note that the non-diagonality is very weak in the case of reactions considered here, and will be finally neglected (see below). Thus the set of equations \eqct{part_shr2} becomes uncoupled. In the numerical procedure we reduce each integro-differential equation \eqct{part_shr2} to a set of linear algebraic equations applying the finite difference and the Simpson methods to approximate the second derivative and the integral, respectively.

\subsubsection{Cluster-folding potential}\label{folding1}
The projectile-target interaction $\hat V_{\vc R,\xi}$ is chosen as a sum of effective complex cluster-target potentials. Each potential is supposed to be a sum of Coulomb, nuclear and spin-orbit terms. The cluster-folding potential $U^{(1)}$ can be defined as the integral
%+++++++++++++++++++++++++++++++++++++++++++++++++++++++++++++++++++++++++++++++++++++++
\begin{eqnarray}
& U^{(1)}_{LL'}(R) = \int d\xi d\Omega_R\, \Theta^{\dag\,L\pr j_0}_{JM}(\xi,\Omega_R) V(\vc R,\xi) \Theta^{Lj_0}_{JM}(\xi,\Omega_R).\nonumber \\ & \label{central2}
\end{eqnarray}
%+++++++++++++++++++++++++++++++++++++++++++++++++++++++++++++++++++++++++++++++++++++++

The cluster model has been used intensively in studies of reactions involving light nuclei. Thus an explicit expression for the cluster-folding potential in some particular cases can be found elsewhere \cite{watanabe, nishioka84}. Therefore we give here its final expression only. Using the Fourier-Bessel transform of the potential $\hat V_{\vc R,\xi}$, and performing the integration over angles and the summation over the momenta projections, we obtain the cluster-folding potential in the following form
%+++++++++++++++++++++++++++++++++++++++++++++++++++++++++++++++++++++++++++++++++++++++
\begin{eqnarray}
& U^{(1)}_{LL'}(R) = \nonumber \\ & \frac1{2\pi^2}  \suml_\lambda (-)^\lambda \hat{L}' \hat \lambda^2 C_{L' 0\lambda0}^{L0} W(j_0J\lambda L;L'j_0) \mc F_{j_0\lambda j_0}(R), \label{FoldingGeneral}
\end{eqnarray}
%+++++++++++++++++++++++++++++++++++++++++++++++++++++++++++++++++++++++++++++++++++++++
where $\hat a = (2a+1)^{1/2}$, $C_{a\alpha b\beta}^{c\gamma}$ is a Clebsch-Gordan coefficient, $W(abcd;ef)$ is a Racah coefficient. The radial factor in \eqrf{FoldingGeneral} reads
%+++++++++++++++++++++++++++++++++++++++++++++++++++++++++++++++++++++++++++++++++++++++
\begin{equation}
\mc F_{j_0\lambda j_0}(R) = \suml_{n=1}^N \suml_{\gamma_0\gamma_0'} \intl_0^\infty q^2 j_\lambda(qR) \tilde v_n(q) \rmtr{j_0\gamma_0'}{C_{n,\lambda}}{j_0\gamma_0} dq, \label{IntegralFactor}
\end{equation}
%+++++++++++++++++++++++++++++++++++++++++++++++++++++++++++++++++++++++++++++++++++++++
where $N$ is a number of clusters, $\tilde v_n(q)$ is a Fourier transform of the cluster-target potential, and operator $C_{n,\lambda\mu}$ is defined as %--------------------------------------------------------------------------------------
\begin{equation}
C_{n,\lambda\mu} = \hat\lambda^{-1} \sqrt{4\pi}\; j_\lambda(qr_n) Y_{\lambda\mu}(\Omega_n). \label{Clm}
\end{equation}
%--------------------------------------------------------------------------------------
The reduced matrix element arising in \eqrf{IntegralFactor} is a partial component of elastic form-factor of $n$-th projectile cluster (i.e. the Fourier transform of its spatial distribution).

The contribution of the spin-orbit interactions $v_i^{(SO)}(r)$ to the folding potential can be calculated in the same way as the central one. However both phenomenological and theoretical analyses \cite{petrovich, amakawa76} show that the spin-orbit interaction plays a minor role even for deuteron elastic scattering and becomes almost negligible in the case of heavy ion elastic scattering \cite{amakawa76}. It was shown \cite{watanabe} that the nucleon-nucleus spin-orbit interaction in the first-order perturbation also gives only a spin-orbit term for the potential of $s$-wave projectile (such as deuteron or $^6$Li). It allows us to consider the folding spin-orbit potential within the approximation proposed in \cite{watanabe, testoni, amakawa76}.

%%------------------------------------------------------------------------------------------------------------------
\subsubsection{Dynamical polarization potential}\label{polariz_2bd}
The calculation of the polarization potential $U^{(2)}(R,R\pr)$ requires the definition of matrix elements of the many-body Green operator $\hat G(z) = (z - \ham_{\xi} - \hat T_{\vc R} - \hat V_{\vc R,\xi})^{-1}$. We write the total Green operator in the form of the Born series as follows
%%----------------------------------------------------------------------------------------------
\eqt{green_lippman}
{\hat G = \hat {\mc G} + \hat {\mc G}(\hat V_{{\vc R},\xi} - \hat {\mc V}) \hat {\mc G} + \dots,}
%----------------------------------------------------------------------------------------------
where the operator $\hat {\mc G} = (z - \hat T_{\vc R} - \ham_{\xi} - \hat {\mc V})^{-1}$ may be factorized if the potential $\mc V$ depends on the relative projectile-target coordinates $\vc R$ only. The appropriate choice of the potential $\mc V(R)$ is the cluster folding potential $U^{(1)}(R)$ obtained above. Note, that in this case $\ham_{\vc R}$ is a nonhermitian operator because the folding potential $U^{(1)}(R) = \mc U(R) + i\mc W(R)$ is complex.

For the total Green operator we use the approximation $\hat G \approx \hat {\mc G}$. By this we neglect the transfer channels, whose contribution is included implicitly through the effective cluster-target interactions. Omitting the second term we neglect also the contribution from the multi-step processes like continuum-to-continuum excitations. The role of continuum-to-continuum coupling has been investigated, in particular, in the paper of Sakuragi et al. \cite{sakuragi87} within the CDCC approach. Significant contribution of the continuum-to-continuum coupling was shown. However, one should point out differences in the formulations of the models in our work and in \rf{sakuragi87}. The authors of \rf{sakuragi87} defined the projectile-target interaction by using the M3Y-type effective nucleon-nucleon interaction with complex normalizing coefficient and applied a double-folding procedure for the calculation of coupling matrix elements. The DPP obtained in \rf{sakuragi87} has an additional large repulsion of the real part in the peripheral region ($Re\; U^{(2)}\sim$ 20 MeV) and an almost negligible additional contribution to the imaginary part. On the other hand, in \rf{makintosh82} the DPP for the similar reaction was calculated within the adiabatic breakup model \cite{amakawa79}, which is quite close to our approach. The DPP found in \rf{makintosh82} provides weak real ($Re\; U^{(2)}<$ 3 MeV) and strong absorptive ($Im\; U^{(2)}\sim -10$ MeV) contributions. Our results are similar.

In \rf{soper} the continuum-to-continuum coupling was also neglected. The authors concluded that this approximation should be valid if the elastic component of the total wave function is larger than break-up ones. Obviously this condition becomes stronger with higher collision energies.

It is difficult to define an explicit applicability condition for our approximation. We need to compare the collision time $t_{coll}$ with the excitation-deexcitation time which is unknown. Instead, we may use the time $t_{int}$ associated with the cluster-cluster relative motion inside the projectile. This gives us the validity criteria in the following form $\tau = t_{coll}/t_{int} \leq 1$. For all reactions considered in this work this condition is fulfilled.

Since the total Hamiltonian $\ham_{\vc R,\xi} \approx \hat T_{\vc R} + \hat {\mc V}_{\vc R} + \ham_{\xi}$ is separable, one constructs a basis as the direct product of the $\ham_{\vc R}$ and $\ham_{\xi}$ bases. The Green operator $\hat {\mc G}$ may be then decomposed over the few-body partial states $\ket{E_p\ve_kJMLj\gamma}{}$ which reads
%+++++++++++++++++++++++++++++++++++++++++++++++++++++++++++++++++++++++++++++++++++++++
\begin{eqnarray}
& \sclr{\xi,\vc R}{E_p\ve_\kappa JMLj\gamma} = \mc C_\xi i^{L} e^{i\sigma_L} \Bigl(\frac2\pi \frac{mp}{\hbar^2}\Bigr)^{1/2}\; \nonumber \\ & \times \frac1{pR}\psi_{JLj}(p,R) \; \Xi_{JM}^{L(j\gamma)} (\kappa,\xi,\Omega_R),\label{partial2bd}
\end{eqnarray}
%+++++++++++++++++++++++++++++++++++++++++++++++++++++++++++++++++++++++++++++++++++++++
where $\ve_\kappa$ is an energy of the cluster-cluster relative motions, and $\mc C_\xi$ is a phase-volume quantity, which depends on the model used for the description of the projectile structure. The spin-angular wave function
%---------------------------------------------------------------------------------------
$$\Xi_{JM}^{L(j\gamma)} (\kappa,\xi,\Omega_R) = \phi_{j\gamma}(\kappa,\xi) \Bigl[ \mc Y_j^\gamma(\Omega_\xi) \otimes Y_L(\Omega_R) \Bigr]_{JM}$$
%---------------------------------------------------------------------------------------
has the same structure as function $\Theta_{JM}^{Lj_0} (\xi,\Omega_R)$ except the sum over $\gamma$.

The dynamical polarization potential is written as
%+++++++++++++++++++++++++++++++++++++++++++++++++++++++++++++++++++++++++++++++++++++++
\begin{eqnarray}
& U^{(2)}_{\nu\nu'}(R\pr,R) = \suml_{\nu_i} \intl_0^\infty |\mc C_\xi|^2 \nonumber \\ & \times \mathbb{V}^\dag_{\nu\nu_i} (\kappa,R\pr)\, g^{(+)}_{\nu_i} (E-\ve_\kappa; R\pr,R) \, \mathbb{V}_{\nu_i\nu'}(\kappa,R) d\ve_\kappa, \label{dpp1}
\end{eqnarray}
%+++++++++++++++++++++++++++++++++++++++++++++++++++++++++++++++++++++++++++++++++++++++
where the matrix elements read
%+++++++++++++++++++++++++++++++++++++++++++++++++++++++++++++++++++++++++++++++++++++++
\begin{eqnarray}
& \mathbb{V}_{\nu\nu_i}(\kappa,R) = \delta_{JJ_i} \delta_{MM_i} \nonumber \\ & \times \int \Theta_{JM}^{\dag\,Lj_0} (\xi,\Omega_R)\; V(\xi,\vc R)\; \Xi_{J_iM_i}^{L_i(j_i\gamma_i)} (\kappa,\xi,\Omega_R)\, d\xi\, d\Omega_R,\nonumber \\ & \label{phi_factor}
\end{eqnarray}
%+++++++++++++++++++++++++++++++++++++++++++++++++++++++++++++++++++++++++++++++++++++++
and the quantum number sets $\nu = \{JML j_0\}$ and $\nu_i = \{J_iM_iL_ij_i\gamma_i\}$. The partial two-body Green function \cite{coulter77} in \eqrf{dpp1}  is
%+++++++++++++++++++++++++++++++++++++++++++++++++++++++++++++++++++++++++++++++++++++++
\begin{equation}
g_{\nu_i}^{(+)}(E_p;R,R')= -\frac{2m}{\hbar^2} \frac1{RR\pr}\;
\frac{\psi_{\nu_i}(p,R_<) \widetilde h_{\nu_i}^{(+)}(p,R_>)}{pS_{\nu_i}} , \label{GreenJL}
\end{equation}
%+++++++++++++++++++++++++++++++++++++++++++++++++++++++++++++++++++++++++++++++++++++++
where notations $R_<$ and $R_>$ refer to the smallest and largest of coordinates $R$ and $R'$. The wave function $\tilde h_{\nu_i}^{(+)}(p,R) = \vf_{\nu_i}(p,R) + i \psi_{\nu_i}(p,R)$ is a combination of two linear-independent solutions of the two-body \shr equation with potential ${\mc V}(\vc R) = U^{(1)}(\vc R)$. The regular wave function $\psi_{\nu_i}(p,R)$ has the boundary conditions \eqct{boundary}, while the irregular solution $\vf_{\nu_i}(p,R)$ has the following asymptotic form
%+++++++++++++++++++++++++++++++++++++++++++++++++++++++++++++++++++++++++++++++++++++++
\begin{eqnarray}
\vf_{\nu_i}(p,R\to 0)  &\sim & (pR)^{-l}, \nonumber \\
\vf_{\nu_i}(p,R\ge R_m) &= & G_L(pR_m) + \frac{S_{\nu_i}-1}{2} H_L^{(+)}(pR_m). \nonumber
\end{eqnarray}
%+++++++++++++++++++++++++++++++++++++++++++++++++++++++++++++++++++++++++++++++++++++++
The correct calculation of the Green function is important for the GOP calculation, since it defines the radial dependance of the polarization potential. Therefore the free-particle Green function, which was used in \cite{testoni, zagrebaev78}, provides a crude approximation and could be used for a qualitative analysis only. Note also, that the local plane wave approximation for the Green function \cite{faessler} gives simple and fast calculation method, which is however less accurate than \eqrf{GreenJL}, but still applicable.

The matrix elements $\mathbb{V}_{\nu\nu_i}(k,R)$ have the same structure as the integral \eqct{central2} except the difference between functions $\Theta$ and $\Xi$. Following the scheme used in the case of the cluster-folding potential, one gets the polarization potential as
%+++++++++++++++++++++++++++++++++++++++++++++++++++++++++++++++++++++++++++++++++++++++
\begin{widetext}
\begin{eqnarray}
U^{(2)}_{L'L}(R\pr,R) & = & \frac1{4\pi^4} \suml_{L_i j_i} \suml_{\lambda'\lambda} (-)^{\lambda'+\lambda} \hat \lambda'^2 \hat \lambda^2 \hat {L}' \hat {L}_i C_{L'0\lambda'0}^{L_i0} C_{ L_i0\lambda0}^{L0} \; W(j_0J\lambda'L_i;L'j_i) \; W(j_iJ\lambda L;L_ij_0) \nonumber
\\ &\times& \suml_{\gamma_0'\gamma_i\gamma_0}\intl_0^\infty |\mc C_\xi|^2 \; \mc F_{(j_0\gamma_0')\lambda'(j_i\gamma_i)}(\ve_\kappa,R')\; g_{JL_i}^{(+)}(E-\ve_\kappa;R',R)\; \mc F_{(j_i\gamma_i)\lambda(j_0\gamma_0)}(\ve_\kappa,R)\; d\ve_\kappa. \label{DPP_2}
\end{eqnarray}
\end{widetext}
%+++++++++++++++++++++++++++++++++++++++++++++++++++++++++++++++++++++++++++++++++++++++
where the functions $\mc F(\ve_\kappa,R)$ read
%+++++++++++++++++++++++++++++++++++++++++++++++++++++++++++++++++++++++++++++++++++++++
\begin{eqnarray}
&\mc F_{(j_0\gamma_0)\lambda(j_i\gamma_i)}(\ve_\kappa,R) = \nonumber \\ &\suml_{n=1}^N \intl_0^\infty q^2 \; j_\lambda(qR) \; \tilde v_n(q)\; \rmtr{j_0\gamma_0}{C_{n,\lambda}}{\ve_\kappa,j_i\gamma_i} dq. \label{TFactor}
\end{eqnarray}
%+++++++++++++++++++++++++++++++++++++++++++++++++++++++++++++++++++++++++++++++++++++++
The reduced matrix element in \eqrf{TFactor} is a component of the transition form-factor, which depends on the projectile structure only and may be calculated once. The integrand in \eqct{TFactor} is an oscillating and decreasing function of $q$, and the integration can be done quite easily with a truncation at $q_{max} \sim 6$ fm$^{-1}$. The resulting function $\mc F(\ve_\kappa,R)$ describes the transition probability and decreases with increasing $\ve_\kappa$. Therefore an integration in \eqrf{DPP_2} may be performed up to some appropriate projectile excitation energy $\ve_k$, which is about 40 MeV in the deuteron case \cite{rawitscher}.

The polarization potential \eqct{DPP_2} is non-diagonal on $L$ and $L'$ indexes. Consequently, one needs to solve numerically the system of coupled integro-differential equations \eqct{part_shr2}. It can be done by an iteration procedure using a solution of the uncoupled system as an initial approximation. However, we avoid this complicated procedure here. The potential \eqct{DPP_2} has a more simple form in the case of the $s$-wave projectile (like deuteron or $^6$Li). It is still non-diagonal ($L' = L,L \pm 2,\dots,L \pm 2j_0$), and this non-diagonality is arising from the non-central part of the cluster-cluster interaction. The calculations show that the non-diagonal terms are about $10^2$ times smaller than the diagonal one, and therefore can be neglected.

\subsubsection{Two-cluster projectile's form-factor}
Let us define the bound and continuum state wave function treating the projectile as a two cluster system (e.g. the deuteron d=p+n). Vector $\vc r = \vc r_1 - \vc r_2$ is an appropriate choice of the coordinate $\xi$ for the description of the cluster-cluster dynamics. A wave function describing bound and scattering states of a two-body system $\ket{\ve,jmls}{}$ ($\gamma \equiv ls$) at a relative energy $\ve$ can be expressed as
%+++++++++++++++++++++++++++++++++++ Equation ++++++++++++++++++++++++++++++
\begin{equation}
\sclr{\vc r}{\ve,jmls} = r^{-1}\; \phi_{jl}(\ve,r) \, \left[ Y_l(\Omega_r) \otimes \chi_s \right]_{jm}, \label{wf_2bd}
\end{equation}
%+++++++++++++++++++++++++++++++++++++++++++++++++++++++++++++++++++++++++++++++++++++++
where the wave function $\phi_{jl}(\ve,r)$ has usual asymptotic at $r\to \infty$. It is either a condition similar to \eqrf{boundary} for a scattering state or the usual asymptotic with the Whittaker function for a normalizable state (see elsewhere). Then using the Wigner-Ekkart theorem, the reduced matrix element of the $C_{n,\lambda\mu}$ operator may be written as follows
%---------------------------------------------------------------------------------------
\begin{eqnarray}
& \rmtr{\ve_\kappa,jls}{C_{n,\lambda}}{j_0l_0s_0} = \delta_{ss_0}\; \mc P_{n,\lambda} (-)^{j_0+j + l_0+l} \hat j \,\hat j_0\; \hat l_0 \; \nonumber \\ & \times C_{l_00\lambda0}^{l0}\; W(l_0s\lambda j;j_0l)\; \tilde \rho_{(jl)\lambda(j_0l_0)}^{(n)}(\ve_\kappa,q), \label{rme_2body}
\end{eqnarray}
%---------------------------------------------------------------------------------------
where quantity $\mc P_{n,\lambda}$ is defined by the symmetry properties of the spherical harmonics $Y_{\lambda\mu}(\Omega_n)$. In particular, $\mc P_{1,\lambda} = 1$ and $\mc P_{2,\lambda} = (-)^\lambda$, since $\vc r_1 = \frac{m_2}{m_1+m_2} \vc r$ and $\vc r_2 = -\frac{m_1}{m_1+m_2}\vc r$, respectively. The function $\tilde \rho^{(n)}(q)$ reads
%---------------------------------------------------------------------------------------
\begin{equation}
\tilde \rho_{(jl)\lambda(j_0l_0)}^{(n)}(\ve_\kappa,q) = \intl_0^\infty \phi_{jl}^*(\ve_\kappa,r)\, j_\lambda(qr_n)\, \phi_{j_0l_0}(r) dr. \label{2BodyTrans}
\end{equation}
%---------------------------------------------------------------------------------------
The quantity $\mc C_\xi$ arising in \eqct{dpp1} in the case of two-cluster system reads $$\mc C_\xi = i^l e^{i\sigma_l} \Bigl( \frac2\pi \frac\mu{\hbar^2\kappa} \Bigr)^{1/2}.$$

%------------------------------------------------------------------------------------------------------
\begin{table*}\caption{\label{OMPTab} Effective interaction potentials.}
\begin{ruledtabular}
\begin{tabular}{rccccccccccc}
&$V_0$\ft{1}&$R_v$ &$a_v$ &$W_0$\ft{2} &$W_D$ &$R_w$ &$a_w$ &$V_{SO}$\ft{3} &$R_{SO}$ &$a_{SO}$ &$R_C$ \\%
\hline %
$       n+p$\ft{4} &62.105& 1.625&      &      &      &      &      &         &      &         &      \\%

$       n+{^{12}C}$& 52.25& 2.57 & 0.570&      & 8.05 & 2.57 & 0.500& 6.2     & 2.29 & 0.750   &      \\%
$       {^{58}Ni}$ & 42.67& 4.53 & 0.750& 7.24 & 2.586& 4.88 & 0.580& 6.2     & 4.26 & 0.750   &      \\%
$      {^{120}Sn}$ & 38.70& 5.77 & 0.750& 7.79 & 0.375& 6.22 & 0.580& 5.5     & 5.18 & 0.750   &      \\%
$      {^{208}Pb}$ & 33.38& 6.93 & 0.750& 6.80 &      & 7.48 & 0.580& 6.2     & 6.52 & 0.750   &      \\%

$     p+{^{12}C}$  & 53.29& 2.57 & 0.570&      & 8.05 & 2.57 & 0.500& 6.2     & 2.29 & 0.750   & 2.75 \\%
$       {^{58}Ni}$ & 44.92& 4.53 & 0.750& 6.10 & 2.214& 5.11 & 0.534& 6.2     & 3.91 & 0.750   & 4.53 \\%
$      {^{120}Sn}$ & 48.46& 5.77 & 0.750& 6.65 & 3.175& 6.51 & 0.627& 5.5     & 5.18 & 0.750   & 5.77 \\%
$      {^{208}Pb}$ & 48.66& 6.94 & 0.750& 8.23 &      & 8.68 & 0.580& 6.2     & 6.52 & 0.750   & 6.93 \\%

$     d+ {^{12}C}$ & 94.65& 2.24 & 0.800& 3.50 & 7.40 & 3.19 & 0.700&         &      &         & 3.29 \\%
%--------------------------------------------------------------------------------
\end{tabular}
\end{ruledtabular}
\footnotetext[1]{Real part has the Woods-Saxon form $V(r) = -V_0\,f(r,R_v,a_v)$, where function $f(r,R,a)=(1+e^{(r-R)/a})^{-1}$.}
\footnotetext[2]{Imaginary part is chosen in the form $W(r) = -W_0\,f(r,R_w,a_w) + 4a_w W_D \frac{d}{dr} f(r,R_w,a_w)$.}
\footnotetext[3]{Spin-orbit interaction has the Thomas form $V_{SO}(r) = 2\lambda_\pi({\bf L\cdot s})V_{SO}\frac1r\frac{d}{dr}f(r,R_{SO},a_{SO})$.}
\footnotetext[4]{The proton-neutron interaction has the Gaussian form $V(r) = -V_0 e^{-r^2/R_v^2}$.}
\end{table*}
%--------------------------------------------------------------------------------

%+++++++++++++++++++++++++++++++++++++++++++++++++++++++++++++++++++++++++++++++++++++++
\section{Application to deuteron elastic scattering}\label{deutron_result}
The deuteron is a two-body projectile with only one bound state. We analyze the elastic scattering of $^2$H as a test of the model. Only the s-wave component of the ground state wave function is considered. The proton-neutron interaction is chosen in a Gaussian form (see Table \ref{OMPTab}) with parameters which give the deuteron binding energy 2.22 MeV, r.m.s. radius $\langle r^2_d \rangle^{1/2} = 1.97$ fm, and triplet scattering length $a_t = 5.46$ fm, close to the experimental data.

As already mentioned, the effect of the deuteron break-up on elastic deuteron-nucleus scattering was analyzed in many papers \cite{testoni, rawitscher, baumg, zagrebaev78, watanabe, perey67, austern87}. We confirm here, in particular, that the break-up transition matrix elements $\mc F_{(j_0l_0)\lambda(jl)}(k,R)$ \eqct{TFactor} with even $\lambda$ dominate in deuteron induced reactions. Odd partial waves are almost negligible at energy well above the Coulomb barrier due to the $\mc P_{2,\lambda}$ coefficient in \eqrf{rme_2body} and the similarity of the p$-$A and n$-$A interactions. The main contributions to the polarization potential come from the transitions with transfer angular momenta $\lambda = 0,2$. The deuteron break-up energies up to about 15 MeV certainly play a leading role in the expansion of the total wave function. It allows one to truncate the sum over $\lambda$ in the polarization potential \eqct{DPP_2} at $\lambda_{max} = 6$ and to perform an integration over the excitation energy $\ve_\kappa$ up to 40 MeV.

In spite of the number of applications of the generalized optical model to the deuteron elastic scattering, a detailed study of the subject has not been reported. In this section we present the results obtained within the approach described above to the deuteron elastic scattering at energies of 30--50 MeV/u. We consider reactions with heavy and light targets separately because of specific features observed in the case of deuteron elastic scattering by light ions.

%+++++++++++++++++++++++++++++++++++++++++++++++++++++++++++++++++++++++++++++++++++++++
\subsection{Scattering by heavy targets}\label{heavy}
The cluster-target potentials for each reaction were obtained from a phenomenological optical model analysis of appropriate experimental data on the n$-$A and p$-$A elastic scattering at energies around $E_{n,p} = E_d/2$ (see discussion in \rf{rawitscher}). We use the optical potential parametrization given in \rf{perey74} in the fitting procedure as a starting parameter set. The obtained parameters are listed in Table \ref{OMPTab}.

The cluster-folding potential $U^{(1)}$ presents well known properties independent of the target \cite{perey67}. It is close to the sum of the nucleon-target optical potentials. The resulting radial dependence of the real part has a Woods-Saxon shape with a noticeably larger diffuseness parameter ($a \sim 1$ fm) and somewhat reduced in magnitude. The same is valid for the imaginary part of $U^{(1)}$. The folding of the spin-orbit nucleon-target interactions, which are usually chosen as a Thomas form (see Table \ref{OMPTab}), results in a shape close to the Woods-Saxon form with parameters similar to the central part.

\begin{figure}
\includegraphics[width=6cm]{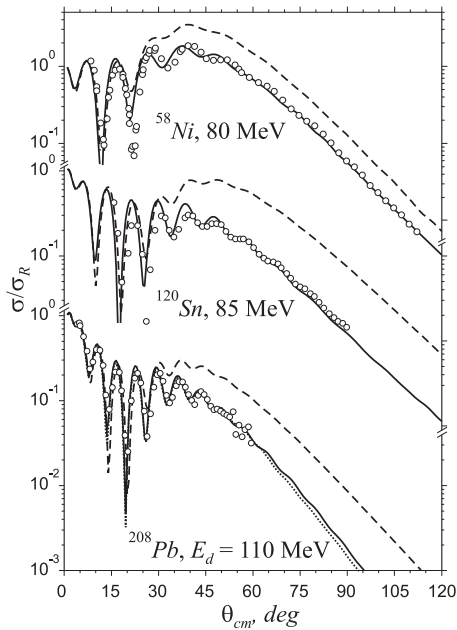}
\caption{Cross sections for deuteron elastic scattering on different targets. Dashed lines show the cross sections obtained with the cluster-folding potential, while solid lines correspond to the calculation with the non-local optical potential. The experimental data from \rf{Steph83, Bojowald88, Betker93} are shown by dots.} \label{A_dd}
\end{figure}

The elastic scattering cross section obtained with the cluster-folding potential $U^{(1)}$ (dashed curves in \fig{A_dd}) significantly overestimates the experimental data at scattering angles larger than the nuclear-rainbow angle. It means that the imaginary part of the folding is too weak. This gives us an estimate of the role of the polarization potential $U^{(2)}$.

%+++++++++++++++++++++++++++++++++++++++++++++++++++++++++++++++++++++++++++++++++++++++
%+++++++++++++++++++++++++++++++++++++++++++++++++++++++++++++++++++++++++++++++++++++++

We calculate the generalized optical potentials $U^{(1)}+U^{(2)}$ for deuteron elastic scattering by  $^{56}$Ni, $^{120}$Sn and $^{208}$Pb targets at energies 80, 85 and 110 MeV, respectively. The corresponding theoretical cross sections are in good agreement with the experimental data (see \fig{A_dd}).
%+++++++++++++++++++++++++++++++++++++++++++++++++++++++++++++++++++++++++++++++++++++++
\begin{figure}
\includegraphics[width=6cm]{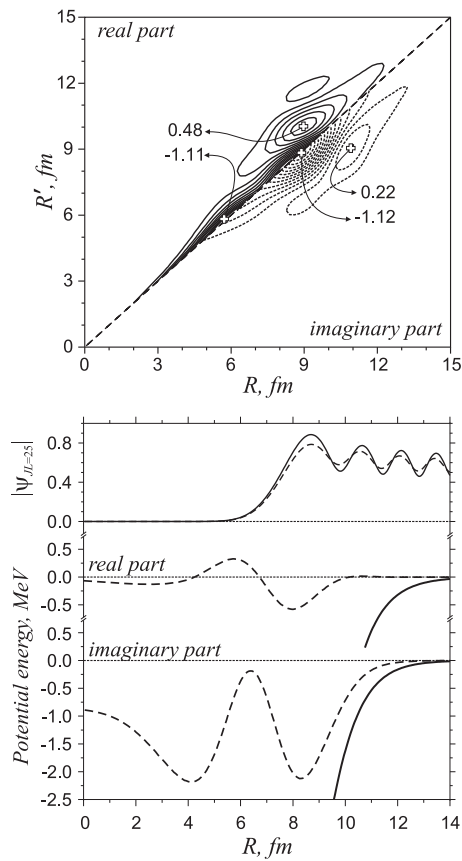}
\caption{Top panel: Dynamical polarization potential calculated for the deuteron elastic scattering on the $^{208}$Pb target at $E_d=$110 MeV and $J$=$L$=25. The real (imaginary) component of the DPP is shown by the solid (dash) contours. The values of the potential at the extremum are indicated. Bottom panel: the module of the partial wave function $\psi_{JL=25}(k,R)$ corresponding to the d(110 MeV) + $^{208}$Pb reaction is shown in top. Solid and dash curves correspond to the wave functions calculated with the $U^{(1)}$ cluster-folding  and with $(U^{(1)}+U^{(2)})$ non-local optical potentials. Real and imaginary parts of the folding potential (solid curves) and WLP (dash curves) is shown for the same reaction in middle and bottom part.} \label{d_DPP}
\end{figure}
%+++++++++++++++++++++++++++++++++++++++++++++++++++++++++++++++++++++++++++++++++++++++

The polarization potential $U^{(2)}_{JL}(R,R')$ is illustrated in the top panel of \fig{d_DPP} for the $^{208}$Pb(d,d) reaction and $J=L=25$. The DPP is symmetric with respect to $R$ and $R'$, therefore we plot the real and imaginary parts of DPP in the same figure. The polarization potential is noticeably non-local, non-monotonic and also $L$-dependent. It makes the polarization potential $U^{(2)}_{JL}(R,R')$ complicated for analysis. The $L$-independent "weighted mean" local polarization potential (WLP) was proposed in \cite{thompson89} as an alternative for the non-local DPP. In the bottom panel of \fig{d_DPP} the WLP for the same reaction is shown together with the module of the corresponding partial wave functions. The wave functions $\psi_{L=25}$ calculated with the non-local DPP and with the WLP are indistinguishable and are shown by the dashed curve in the top part of the bottom panel of \fig{d_DPP}. Note also that the cross sections obtained with the WLP and with the initial non-local DPP are almost identical. It is illustrated in \fig{A_dd} for the case of lead target, where the dotted line shows the cross section obtained with the WLP. It allows us to analyze the properties of the WLP instead of the DPP.

The relative contribution of polarization potential to the real part of the optical potential is rather small. In particular, the value of $\mc{Re}\,U^{(2)}_{WLP}(R)$ at the minimum around $R$ = 8 fm is about -0.7 MeV and amounts to 3\% of the folding potential $U^{(1)}$, whereas the contribution of the imaginary part of the polarization potential is more than 30\% of the folding potential at this point. For the d + $^{208}$Pb  collision the total reaction cross section is $\sigma_R$ = 2.9 b at $E_d$ = 110 MeV, where the folding potential gives $\sigma_R^{(F)}$ = 2.73 b and the polarization potential gives $\sigma_R^{(DPP)}$ = 0.17 b.

Thus, we may conclude that the generalized optical potential model provides an adequate description of the light two-cluster projectile elastic scattering by heavy nuclei at intermediate energies. The deuteron polarization due to the coupling with the break-up channels properly describes the missing part of the total reaction cross section (about 10\%).
%+++++++++++++++++++++++++++++++++++++++++++++++++++++++++++++++++++++++++++++++++++++++
\begin{figure}
\includegraphics[width=8cm]{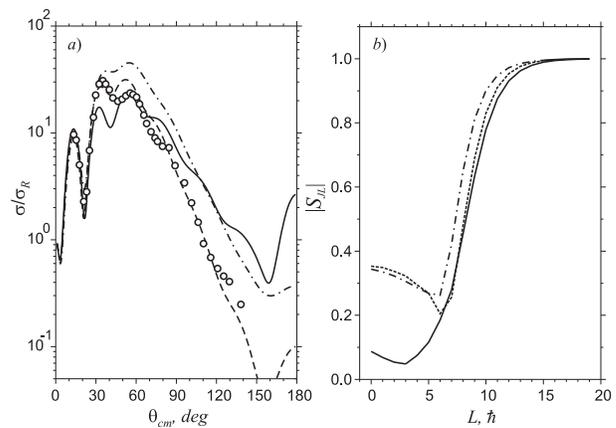}
\caption{Elastic scattering cross sections ($a$) and $S$-matrices ($b$) for the $^{12}C(d,d)$ reaction at 56 MeV. Dash-dotted and solid curves in both panels correspond to the results obtained with cluster-folding and non-local optical potentials. The long dashed line in panel ($a$) shows the cross section obtained by neglecting the polarization potential at $L<6$. The short dashed line in panel ($b$) is an $S$-matrix generated by phenomenological optical potential. Dots are the experimental data taken from \cite{Matsuoka86}.} \label{d12C56MeV}
\end{figure}
%+++++++++++++++++++++++++++++++++++++++++++++++++++++++++++++++++++++++++++++++++++++++

%+++++++++++++++++++++++++++++++++++++++++++++++++++++++++++++++++++++++++++++++++++++++
\begin{figure*}
\includegraphics[width=12cm]{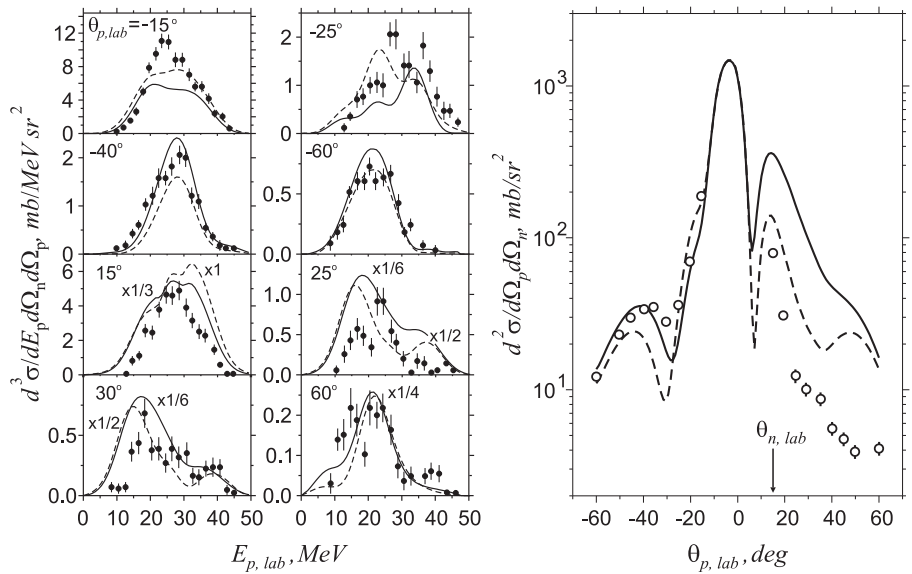}
\caption{Left panel: energy spectra of protons in coincidence with neutrons emitted at $\theta_{n,lab}=15^\circ$ for the $^{12}C(d,pn)$ elastic break-up at 56 MeV in the angular region $-60^\circ \le \theta_{p,lab} \le 60^\circ$. Right panel: angular distribution of $p$-$n$ correlations in the same reaction. Solid and dashed curves are the DWBA calculations of the break-up cross section. Dashed lines show results by omitting the contribution of the projectile-target partial waves with $L<6$. The calculated triple differential cross sections at positive $\theta_{p,lab}$ are renormalized by the factor shown near each curves. Dots are the experimental data from \cite{Matsuoka82}.} \label{C_dpn_56MeV}
\end{figure*}
%+++++++++++++++++++++++++++++++++++++++++++++++++++++++++++++++++++++++++++++++++++++++

%--------------------------------------------------------------------------------------
\subsection{Scattering by light targets}\label{light}
We study here the d + $^{12}$C collision at 56 MeV, because there are experimental data both on the elastic scattering \cite{Matsuoka86} and on the proton-neutron correlations in the $^{12}$C(d,pn) break-up reaction \cite{Matsuoka82}. We use the parameters of the nucleon-carbon potentials taken from \rf{Matsuoka82} (see Table \ref{OMPTab}).

The general properties of the optical potential are the same as in the case of reactions with heavy targets. The elastic scattering cross section calculated with the folding potential exceeds the experimental data, see \fig{d12C56MeV}($a$). However, in contrast with heavy targets an addition of the polarization potential to the folding does not lead to agreement with experimental data. Comparison of the partial $S$-matrix elements generated by the folding (dash-dotted line) and by the non-local optical potential (solid line) shows that the polarization potential provides a strong additional absorption at low orbital momenta.

We perform a fit of experimental data on the d + $^{12}$C elastic scattering within the usual phenomenological optical model \cite{nrvOM} using the cluster-folding interaction as an initial approximation. The fitted OMP parameters (see Table \ref{OMPTab}) provide an angular distribution which agrees with the experimental points. The corresponding $S$-matrix elements are shown in \fig{d12C56MeV}($b$) by the short-dashed curve. As it can be seen, there is a good agreement of the partial $S$-matrix elements obtaining with the generalized and phenomenological optical potentials at $L \ge 6$ and significant differences at $L<6$. Also the phenomenological $S_{JL}$ elements are rather close to those obtained with the folding at low values of $L$, i.e. the break-up probability for central collisions should be small. Thus, one may conclude that the model does not describe properly the deuteron break-up and, consequently, elastic scattering at low partial waves.

\subsubsection{Deuteron break-up within the prior-form DWBA}
The function $\mc F_{(jl)\lambda(j_0l_0)}(k,R)$ in the polarization potential \eqct{DPP_2} determines also the prior-form of the DWBA break-up amplitude \cite{rybicki72} %--------------------------------------------------------------------------------------------
\begin{equation}
T^{DW} = \mtr{\psi_{\vc p'}^{(-)}(\vc R)\, \phi_{\vc k}^{(-)}(\vc r)}{V_{\vc R,\vc r}}{\phi_{g.s.}(\vc r)\, \psi_{\vc p}^{(+)}(\vc R)}, \label{TDWBA}
\end{equation}
%--------------------------------------------------------------------------------------------
where $\phi_{g.s.}(\vc r)$ and $\phi_{\vc k}^{(-)}(\vc r)$ are the ground and excited states of the projectile, and $\psi_{\vc p}^{(\pm)}(\vc R)$ are the distorted waves describing the projectile-target relative motion in the entrance and exit channels. We calculated the deuteron break-up cross section for the $^{12}$C(d,pn) reaction at 56 MeV and compared it with experimental data \cite{Matsuoka82}, where the angular and energy distributions of the protons were measured in coincidence with neutrons emitted at $\theta_{n, lab} = 15^\circ$. Here we used the same OMP parameters as in analysis of d+$^{12}$C elastic scattering above.

The results are shown by solid lines in \fig{C_dpn_56MeV}. The DWBA amplitude \eqct{TDWBA} gives a good agreement with the data for negative proton angles and fails at positive ones. One may suppose that the differences between the calculations and experimental data on deuteron break-up and elastic scattering have the same origin.
%+++++++++++++++++++++++++++++++++++++++++++++++++++++++++++++++++++++++++++++++++++++++
\begin{figure}
\includegraphics[width=7cm]{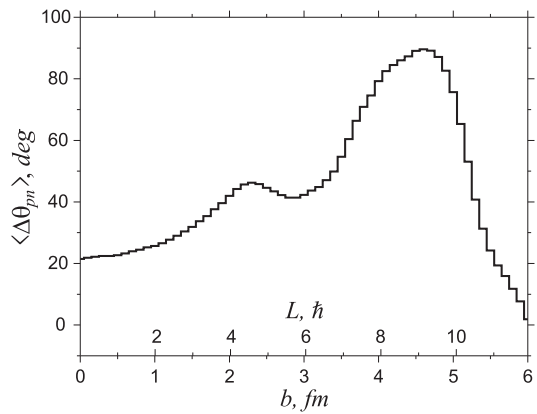}
\caption{Averaged correlation angle between the protons and neutrons emitted in the $^{12}$C(d,pn) break-up reaction at $E_d$ = 56 MeV as a function of the impact parameter. The calculation is performed within the classical three-body model \cite{semkin95}.}
\label{classic}
\end{figure}
%+++++++++++++++++++++++++++++++++++++++++++++++++++++++++++++++++++++++++++++++++++++++

\subsubsection{Deuteron break-up within the classical dynamics model}
To confirm this assumption we performed an analysis of this reaction within the few-body classical molecular dynamics \cite{semkin95}. The models based on the Newtonian equations have been successfully applied to the study of heavy-ion fragmentation at intermediate energies (see, for example, \cite{royer87, wada93, mohring91, semkin95, denikin03}). Note also that classical dynamics approaches turn out to be very effective in combination with quantum consideration, allowing to explain many aspects of nuclear dynamics using a "trajectory" language.

Within the classical model the two-body projectile d=(p+n) and target $^{12}$C are treated as classical particles moving along the classical trajectories determined by the same interactions ($V_{\vc R,\vc r} + v_{\vc r}$) as in the quantum case considered above. We tested $10^6$ trajectories with randomly distributed initial parameters (see details in \cite{semkin95}). \fig{classic} shows the averaged angle $\langle\Delta\theta_{pn}\rangle$ between the proton and neutron emitted during the deuteron break-up as a function of the impact parameter $b$ = $L/p$. As it can be seen, protons emitted at negative angles (relative to $\theta_{n,lab}$ = 15$^\circ$, large $\Delta\theta_{pn}$) are originated mainly in peripheral collisions with $L\ge6$, whereas positive angles $\theta_p$ correspond to central collisions (small $\langle\Delta\theta_{pn} \rangle$ values). Hence, the disagreement with experiment at positive values of $\theta_p$ (\fig{C_dpn_56MeV}) originated from a wrong treatment of the contribution of small angular momenta ($L<6$) -- just as in the analysis of elastic scattering.

Note that a large $\langle\Delta\theta_{pn}\rangle$ value in peripheral deuteron break-up process is caused by the repulsion of protons by the Coulomb field of the target, whereas the neutron is deflected by attractive nuclear forces. In central collisions the effect of the Coulomb forces is much weaker, therefore the $\langle \Delta\theta_{pn} \rangle$ angle turns out to be relatively small.

\begin{figure}
\includegraphics[width=6cm]{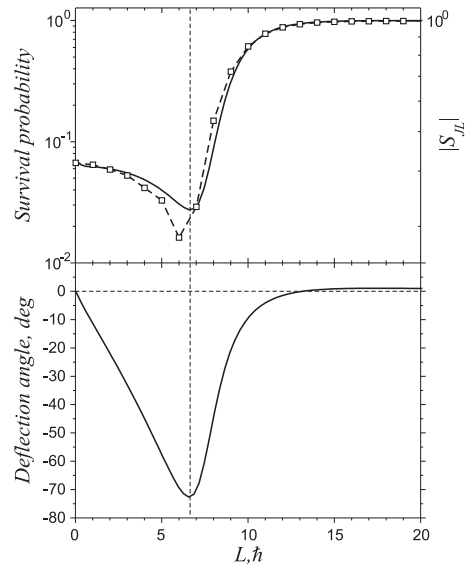}
\caption{Deuteron survival probability and deflection function calculated within classical model \cite{nrv} for the d(56 MeV) + $^{12}$C reaction. Open squares show the $S_{JL}$ matrix obtained within the phenomenological optical model with the same potential.}
\label{SurvDF}
\end{figure}

The origin of the critical orbital momentum $L = 6$ (for d+$^{12}$C at 56 MeV) also has a clear explanation in the classical model. Using the phenomenological optical potential for the d(56 MeV) + ${^{12}}$C reaction (see Table \ref{OMPTab}), and employing an appropriate computational code \cite{nrv} we calculated the classical deflection function and survival probability
%--------------------------------------------------------------------------------------------
\begin{equation}
P_{s}(L)  = \exp\Bigl(-\intl_{tr} \frac{\mc W(R)dR} {\sqrt{\frac{\hbar^2}{2m} [E_p-\mc U(R)]}} \Bigr) \label{Pclas}
\end{equation}
%--------------------------------------------------------------------------------------------
as a function of the orbital momentum $L = b p$. $P_s(L)$ is the probability that the projectile, moving along the trajectory with a given impact parameter $b$, remains in the elastic channel. $\mc U(R)$ and $\mc W(R)$ in \eqct{Pclas} are the real and imaginary parts of the optical potential, and the integration is performed along the trajectory. The calculated survival probability $P_{s}$ are very similar to the partial $S$-matrix elements, which have the same physical meaning (see \fig{SurvDF}). The deflection function reveals the nuclear rainbow angle $\theta_{NR} \approx 70^\circ$ close to the experimental value.
%+++++++++++++++++++++++++++++++++++++++++++++++++++++++++++++++++++++++++++++++++++++++
%+++++++++++++++++++++++++++++++++++++++++++++++++++++++++++++++++++++++++++++++++++++++
As it can be seen from \fig{SurvDF}, the orbital momentum $L_{NR}=6$ corresponds to the nuclear rainbow scattering. This means that the trajectories with $L<L_{NR}$ pass deeply in the interaction region, while the trajectories with $L>L_{NR}$ are more peripheral. Thus the model used here for a calculation of the generalized optical potential does not treat properly central collisions with a strong overlapping of the colliding nuclei.

%+++++++++++++++++++++++++++++++++++++++++++++++++++++++++++++++++++++++++++++++++++++++
\begin{figure*}
\includegraphics[width=13cm]{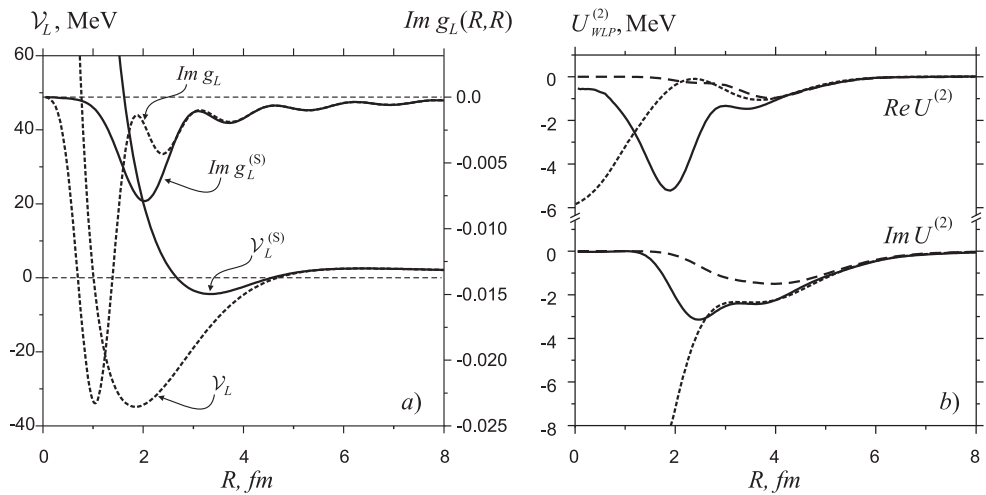}
\caption{($a$) Imaginary part of the partial Green functions ($L$=2) and real part of corresponding interaction potentials for the $^{12}$C(d,d) reaction. Functions $g_L$ and $g_L^{(S)}$ are calculated using effective potential $\mc V_L = U^{(1)} + \hbar^2L(L+1)/2mR^2$ and its supersymmetric partner $\mc V_L^{(S)}$ after removal of the state $E_2=-(18.5+i9.3)$ MeV, respectively. ($b$) WLPs for the same reaction resulted from calculation with non-modified non-local optical potential (dotted line), with non-local potential neglected polarization part at $L<6$ (dash line), and with non-local optical potential after removal forbidden states (solid line).} \label{SUSYGreen}
\end{figure*}
%+++++++++++++++++++++++++++++++++++++++++++++++++++++++++++++++++++++++++++++++++++++++

\subsubsection{Projectile-target non-physical bound states}
The phenomenological OM analysis shows that the absorption at low orbital momenta in the elastic scattering is well described by the imaginary part of the folding potential only and does not require any addition  (compare dash-dotted and dashed curves in \fig{d12C56MeV}($b$)). Note that a simple cut of the polarization potential at $L<6$ in the elastic scattering analysis and omitting the contribution of these partial waves to the DWBA break-up cross-section lead to a significant improvement of the results in both cases. It is shown by the dashed curves in \fig{d12C56MeV}($a$) and \fig{C_dpn_56MeV}. Thus, the experimental data indicates that the deuteron in the inner region of the target nucleus turns out to be stable relatively to the break-up channels, that is confirmed also in previous studies \cite{testoni}.

Thus, we conclude that the calculated DPP overestimates the absorption at small partial waves, i.e. at small projectile-target relative distance. The radial dependence of the polarization potential in this region is defined mainly by the partial Green function $g_{JL}^{(+)} (R,R')$ \eqct{GreenJL}. The properties of the Green function at low $L$ values are significantly affected by the properties of the folding potential $U^{(1)}(R)$ which is used to calculate the Green function. The interaction $U^{(1)}(R)$ is calculated as a sum of the folded complex cluster-target optical potentials. Parameters of these potentials are usually fitted in order to reproduce experimental cross sections. In this procedure the scattering phase-shifts are retrieved but not the wave functions. Thus, the cluster-target potentials may provide an incorrect behavior of the partial wave functions at small distances because of $\pi$ ambiguity of the phases. In particular, cluster-target potentials and, consequently, cluster-folding one turn out to be deep and contain many forbidden bound states. This may results in incorrect radial dependence of the Green function at small distance, since the corresponding partial wave functions penetrate deeply into the interaction region. This leads to the rise of the DPP at small $R$ values.

The observed stability of deuteron moving in nuclear matter with respect to the break-up means in fact that the deuteron does not penetrate deeply into a target due to the Pauli blocking. The effects of antisymmetrization in deuteron elastic scattering have been studied before \cite{PauliBlock}. In order to take it into account consistently within our approach one needs to remove the forbidden states from the nucleon-target potentials, that makes them non-local and results in the complication of their further treatment. Therefore we apply a simplified method modifying the d$-^{12}$C cluster-folding potential $U^{(1)}(R)$ which also presents a number of non-physical bound states. Let us then remove these states.

States found in the d$-^{12}$C folding potential (without spin-orbit interaction) are listed in Table \ref{Levels}. $E'_{nL}$ are the eigenvalues corresponding to the states in the potential without imaginary part, while $E_{nL}$ are the eigenvalues in the complex potential (since the cluster-folding interaction is complex). The imaginary part of the potential leads to the appearance of a negative imaginary addition to the eigenvalues. $\re\;E_{nL}$ of the bound states as well as of the narrow resonances are modified a little, while the broad resonances are shifted significantly.
%--------------------------------------------------------------------------------
\begin{table}
\caption{\label{Levels} Bound, resonant and normalizable states in the $d+{^{12}C}$ folding potential.}
\begin{ruledtabular}
\begin{tabular}{cccc} L & nodes & $E'_{nL}$(MeV) & $E_{nL}$(MeV)\\%
\hline %
 0 & 0 & --51.47 + $i$0     &--51.29 -- $i$8.25 \\%
 1 & 0 & --34.60 + $i$0     &--34.52 -- $i$9.11 \\%
 0 & 1 & --19.72 + $i$0     &--19.57 -- $i$8.34 \\%
 2 & 0 & --18.57 + $i$0     &--18.48 -- $i$9.25 \\%
 1 & 1 &  --6.93 + $i$0     & --6.37 -- $i$6.82 \\%
 3 & 0 &  --4.03 + $i$0     & --3.62 -- $i$8.39 \\%
 0 & 2 &    0.14 -- $i$0.0  &   0.49 -- $i$0.65 \\%
 1 & 2 &    1.24 -- $i$2.10 &   0.82 -- $i$1.25 \\%
 2 & 1 &    1.89 -- $i$0.18 &   1.76 -- $i$1.72 \\%
 3 & 1 &    5.95 -- $i$4.51 &   3.51 -- $i$3.49 \\%
 4 & 0 &    7.36 -- $i$0.59 &   7.11 -- $i$3.57 \\%
 5 & 0 &   16.45 -- $i$4.88 &  12.12 -- $i$7.45 \\%
 0\ft{1} &  &             &    1.47 -- $i$3.06 \\%
 2\ft{1} &  &             &    3.85 -- $i$3.57 \\%
 4\ft{1} &  &             &    9.57 -- $i$5.78 \\%
%--------------------------------------------------------------------------------
\end{tabular}
\end{ruledtabular}
\footnotetext[1]{The normalizable states in the $d+{^{12}C}$ folding potential with positive real energy.}
\end{table}
%--------------------------------------------------------------------------------

For an hermitian Hamiltonian the $S$-matrix poles corresponding to resonant states are symmetric with respect to the imaginary $p$-axis. In the case of a complex potential this symmetry is broken. Generally the resonant poles move in clockwise direction in the complex $p$-plane (see \rf{sparenberg96} for details). The states with $\re\, p > 0$ (right half-plane) get negative addition to the $\im\,p$, while some of the poles in the left complex half-plane cross the real axis and become normalizable states ($\re\,p <0$ and $\im\, p > 0$, i.e. $\psi_L(R\to \infty) \sim e^{-\im\, p R}$). The energies corresponding to these states in the d$-^{12}$C folding potential are also listed in Table \ref{Levels}.
%+++++++++++++++++++++++++++++++++++++++++++++++++++++++++++++++++++++++++++++++++++++++
\begin{figure*}
\includegraphics[width=13cm]{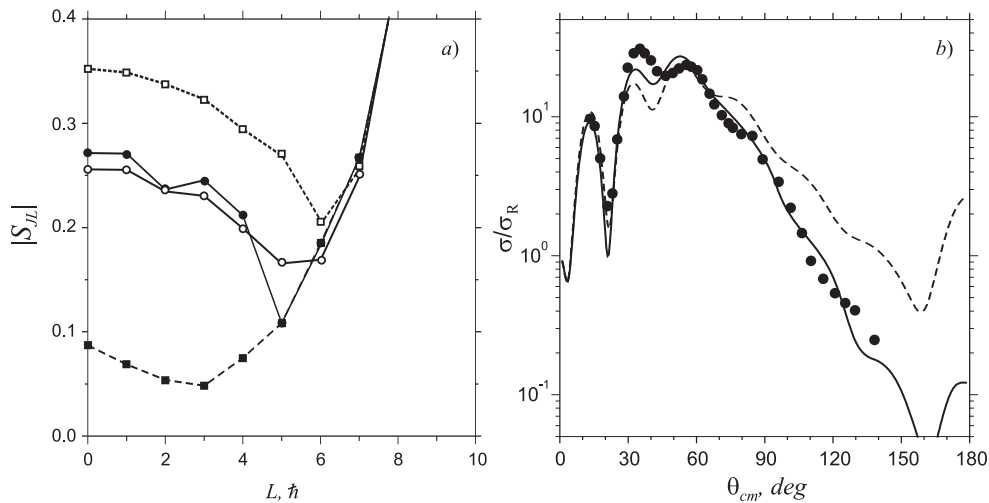}
\caption{The $S$-matrices for $^{12}$C(d,d) reaction at energy $E_d$ = 56 MeV are compared in panel ($a$). Full and open squares show the $S$-matrix generated by non-modified non-local and empirical optical potential, respectively. Full and open circles are the S-matrix elements calculated with non-local optical potential after removal forbidden states and with the corresponding WLP, respectively. ($b$) Elastic scattering cross section for the same reaction. The solid curve shows the calculations with the WLP after removal of the forbidden states, while the dashed curve shows the calculation with non-modified non-local optical potential. Dots are the experimental data \cite{Matsuoka86}.} \label{dC_SUSY}
\end{figure*}
%+++++++++++++++++++++++++++++++++++++++++++++++++++++++++++++++++++++++++++++++++++++++

We apply the technique explained in \cite{baye87,baye96}, which allows to eliminate normalizable states from the spectrum of the complex potential using supersymmetric transforms in each partial wave. The resulting potential becomes $L$-dependent, and contains a strong repulsive core at small distances. The supersymmetric transforms do not modify the scattering phase-shifts. Nevertheless the partial wave function and, consequently, the Green function turns out to be pushed out from the interaction region (see \fig{SUSYGreen}($a$)). The obtained Green function was used in the DPP calculation. The corresponding WLP are shown in \fig{SUSYGreen}($b$) in comparison with the initial WLP and with the WLP resulted from a dropping of the non-local polarization potential at $L<6$. We may conclude that the elimination of the forbidden states modifies the DPP in a correct way.

Modules of the $S$-matrix elements for the same reaction are shown in \fig{dC_SUSY}($a$). One may see that the supersymmetric transform leads to a decrease of the absorption in elastic channels at low partial waves. Thus the elimination of the non-physical states in the d$-^{12}$C folding potential allows one to describe effectively the suppression of the deuteron break-up at low values of angular momenta.

The d$-^{12}$C folding interaction does not contain normalizable states with $L=5$ as it may be expected from the behavior of the phenomenological $S$-matrix (see \fig{dC_SUSY}(a)). This indicates that the folding potential is not the best substitution for the $V_{\vc R, \vc r}$ in the Green function calculation. Damped $S_{L=5}$ matrix element (solid circles in \fig{dC_SUSY}(a)) results in the oscillating behavior of the cross section at large scattering angles. This problem is overcame somehow if we use the WLP (solid line in \fig{SUSYGreen}($b$)), which smoothes the $S$-matrix (open circles in \fig{dC_SUSY}($a$)) by averaging the non-local polarization potential over all orbital momenta. The angular distribution in the d+$^{12}$C elastic scattering is shown in \fig{dC_SUSY}($b$) together with experimental data and with the cross section obtained without the supersymmetric transforms.

%--------------------------------------------------------------------------------------
\section{Conclusions}\label{concl}
By extending the model proposed in earlier papers, we derive the generalized optical potential for elastic scattering of a few-cluster projectile, taking into account explicitly the coupling with the break-up channels. We do not use most of the simplifications which were employed in previous papers. In particular, applying the model to deuteron elastic scattering we take into account the spin of projectile, consider the coupling to the projectile continuum with cluster-cluster relative orbital momenta $l\le6$, and apply the suitable approximation of the Green function instead of the free-particle one used before. It allows to improve an agrement with the data and previous results obtained within different approaches, that supports the efficiency of the model. The model was applied to study of the deuteron elastic scattering at energies of few tens of MeV per nucleon and good agreement with experimental data was obtained.

It was also shown that the behavior of the polarization potential at low orbital momenta is noticeably affected by the non-physical bound states in the projectile-target system. In the case of light heavy-ion scattering it leads to the overestimation of the absorption in the GOP in low partial waves. The elimination of these forbidden states allows one to obtain an appropriate polarization potential.

Note that the non-physical bound states do not reveal itself in the deuteron scattering by heavy targets in spite of their existence for low partial waves. The reason is a much stronger absorption part of the folding potential in the case of heavy targets as compared with light nuclei. Addition of the polarization potential to the folding one gives a negligible effect on the elastic scattering cross section in low partial waves because the contribution of these partial waves is suppressed by the absorptive part of the folding interaction. Note, however, that the forbidden states may play some role in other reaction channels, for example, in break-up. This subject is an interesting problem for future studies.

Application of the proposed model to the reactions with a three-cluster weakly bound nuclei (such as $^6$He = $\alpha$ + n + n) will be done in future works.

\begin{acknowledgments}
This work was supported by the INTAS (Grant No. 04--83--2649) and by a grant of FNRS (Belgium). One of the authors (A.D.) would like to thank the PNTPM department of the Universit\'{e} Libre de Bruxelles for hospitality. Dr. J.-M. Sparenberg is kindly appreciated for useful discussions and providing the SUSY transformation code.
\end{acknowledgments}

%--------------------------------------------------------------------------------

%%--------------------------------------------------------------------------------------
\end{document}